\documentclass[prl,aps,superscriptaddress,twocolumn,floats,nofootinbib,showpacs]{revtex4}
\usepackage{amsmath,amssymb,graphicx}

\begin{document}

\title{Coulomb blockade with neutral modes}

 \author{Alex Kamenev}
\affiliation{William I. Fine Theoretical Physics Institute,
and School of Physics and Astronomy,
University of Minnesota, Minneapolis, MN 55455}

\author{Yuval Gefen}
\affiliation{Department of Condensed Matter Physics, Weizmann Institute of Science, Rehovot, 76100 Israel}

\begin{abstract}
We study transport  through a quantum dot in the fractional quantum Hall regime with filling factors $\nu=2/3$ and $\nu=5/2$, weakly coupled to the leads. We account for 
both injection of electrons to/from the leads, and quasiparticle rearrangement processes between the edge and the bulk of the quantum dot. The presence
of neutral modes introduces topological constraints that modify qualitatively the features of the Coulomb blockade (CB). The periodicity of CB peak spacings doubles and the ratio of spacing between adjacent peaks approaches (in the low temperature and large dot limit) a universal value: $2:1$ for $\nu=2/3$ and $3:1$ for $\nu=5/2$.  The corresponding  CB diamonds alternate their width in the direction of the bias voltage and 
allow for the determination of the neutral mode velocity, and of the topological numbers associated with it.

\end{abstract}

\pacs{ }
\maketitle

Coulomb blockade (CB) conductance oscillations in small quantum Hall islands is
proven to be a useful tool to study the structure of the edge modes. Early experiments
\cite{McEuen,Foxman} performed in the integer quantum Hall effect (QHE) regime were instrumental
to demonstrate the presence of compressible and incompressible stripes along the edge \cite{Shklovskii,Beenakker}.
Various generalizations to tunneling through a quantum dot in the fractional quantum Hall regime have been proposed \cite{Maksym, Kinaret}.

More recent effort has been focused on  the nature and structure of the fractional QHE edges at composite filling
factors such as $\nu=2/3$ \cite{Meir, Beenakker92,Heiblum2010,Yacoby,Heiblum}, $\nu=5/2$  \cite{Heiblum2010,Yacoby-1,Smet,Akhmerov}, and other filling fractions \cite{Hiro}.
Early models of the  $\nu=2/3$  edge \cite{Beenakker92} posited that the current is carried by two  edge channels, each having
the characteristics of a $\nu=1/3$ edge state. Other models \cite{MacDonald} introduced  counter-propagating edge channels corresponding to $\nu=1$ and $\nu=-1/3$.
These models  yield a Hall conductance which is not quantized and is
non-universal, and do not explain certain experimental facts \cite{Wang}. 
Following Wen's \cite{Wen} description of the  $\nu=1/3$ fractional QHE edge Kane, Fisher and Polchinski
considered the composite $\nu =2/3$ edge \cite{KFP}. They showed that the two original counter-propagating modes are strongly mixed by both disorder and inter-mode interaction. Under fairly general conditions such mixing
results in the emergence of a  forward $2/3$ {\em charge}  mode along with a backward moving {\em neutral} mode. The latter
carries  energy (heat), but no electric charge. While neutral modes  have been detected in various setups \cite{Heiblum2010,Yacoby,Hiro},  understanding the full implications of the physics involved, and characterising their topological numbers,  remains a major challenge. This task gained significant urgency
since the structure of the edge modes provides an insight into the possible non-Abelian nature of the $\nu=5/2$ anti-Pfaffian state \cite{MooreRead,ReadGreen}, explored for quantum information purposes \cite{Nayak}.

\begin{figure}
\vspace{-2.5cm}
\hspace{-1cm}
\begin{center}
  \includegraphics[width=9cm]{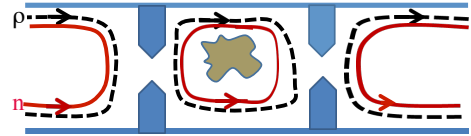}
  \vspace{-3cm}
\caption{(Color online) A quantum dot in $\nu=2/3$ regime. The two renormalized edge channels represent the charge mode (dashed line, black) and the neutral mode (solid line, red). The bulk of the quantum dot is represented by a puddle (green).}
\label{fig1}
\end{center}
\end{figure}

It was suggested \cite{Grosfeld} that  the pattern of the CB peaks as well as  thermopower measurements \cite{Stern} in a QHE island with the composite filling fraction may be used to detect the presence of neutral modes. In particular it has been argued that CB peak spacing exhibits a
slight period doubling in $\nu=2/3$ state \cite{Meir,Stern,Shklovskii}.  A more complicated modulation of CB peak structure was predicted for other fractions \cite{Parsa}, including  $\nu =5/2$ (the structure will then depend on the exact nature of the ground state  \cite{MooreRead,ReadGreen,Lee,Levin}).

The present paper  focuses on the structure of CB {\em diamonds} (i.e. non-linear source-drain conductance) along with the peak spacing
in composite fraction QHE dots. As opposed to previous studies, here we study the direct consequences of quasiparticle exchange between the edge and the bulk of the
quantum dot.  Our main findings are: (i) We show that, under rather general assumptions regarding the electrostatic energies of the dot,
such edge-bulk exchange leads to a distinct universal pattern of CB diamonds as well as peak spacing. In particular, we find that for $\nu=2/3$ the CB peak
spacing  tends to a universal $2:1$ ratio between odd and even states in the limit of large dot. At the same time the Coulomb diamonds are wider in the source-drain voltage directions (i.e. larger gap) between closely spaced peaks and narrower (i.e. smaller gap) between remotely spaced peaks, Fig.~\ref{fig3}. (ii) The width of the Coulomb diamonds in the source-drain bias provides a direct information on the velocity of the neutral modes, and consequently on the topological numbers (so-called zero modes) associated with the latter. (iii) We carry a similar analysis for $\nu=5/2$ systems.  Assuming an anti-Pfaffian state \cite{MooreRead,Levin},  we find a tendency (in the limit of large dot) towards a universal  $3:1$ ratio between odd and even states, and a similar (though distinct) pattern of CB diamonds as well.

Consider an (almost) isolated 2D island of QHE liquid confined by electrostatic gates \cite{McEuen}, Fig.~\ref{fig1}. It contains an integer number of electrons $N$. If the external magnetic field is such that the Landau level filling factor is a fraction, e.g. $\nu=2/3$, the bulk of the dot has a correlated ground state  separated from the excitation spectrum by a gap. The only low energy excitations are those localized along the edge \cite{Wen,KFP}. There may be more than one type of such edge excitations having different charges and velocities \cite{Meir}. For a macroscopic sample all edge modes are gapless, while for a closed dot they acquire a small gap which scales as $ v h/R$, where $v$ is the mode velocity, $h$ is its scaling dimension and $R$ is the radius (or rather circumference$/2\pi$) of the dot. Imagine now varying the gate voltage, thus increasing the effective area of the dot and consequently the number of flux quanta penetrating it.  Provided  the number of electrons, $N$, remains unchanged, this would decrease the filling factor $\nu$. As a result, it either creates fractionally charged quasiparticle (actually quasihole) excitations in the bulk of the dot or on its edge \cite{Grosfeld}.
Which of the two is favorable  depends on the ratio between the charging energy of a quasiparticle in the bulk, and the edge gap mentioned above. An important observation is that this balance depends on the gate voltage and may ubruptly flip upon an adiabatic variation of the latter. This is the mechanism of the quasiparticle exchange between the bulk and the edge at fixed $N$, central to our discussion.

Upon further variation of the gate voltage, it may become energetically favorable to move one electron from the lead to the dot. If two leads are attached, Fig.~\ref{fig1}, a current may flow through the dot under a small source--drain voltage. This is reflected in a sharp rise of the linear conductance at certain gate voltages, known as CB peaks. It is important to notice that it is an {\em electron}, and not a fractionally
charged quasiparticle, which is exchanged between the dot and the leads.  The positions of the CB peaks are sensitive to the energy cost of accommodating an additional electron. The latter depends on the way the electron is decomposed onto fractionally charged edge modes and bulk quasiparticles.  


To be specific, consider a quantum dot in the $\nu=2/3$ QHE regime. If the dot is sufficiently large, disorder and  interactions lead to the formation of charge and neutral modes \cite{KFP}, described by the bosonic fields $\phi_\rho$ and $\phi_{\mathrm n}$ correspondingly (the charge density is given by $\rho_c=\nu^{1/2}\partial_x\phi_\rho/(2\pi)$). Under these conditions, the elementary edge excitations are given by a triplet of quasiparticles and a doublet of electrons, described by the operators \cite{KFP,Stern}
\begin{eqnarray}
&&\Psi^{qp}_{1,\pm} = e^{\pm i\phi_{\mathrm n}/\sqrt{2}}\,  e^{i\phi_\rho/\sqrt{6}}\,, \quad\quad q={1\over 3},\quad h={1\over 3}\,;\nonumber \\
&&\Psi^{qp}_{0,0} =  e^{i2\phi_\rho/\sqrt{6}}\,, \quad\quad q={2\over 3},\quad h={1\over 3}\,; \label{eq:quasiparticles}  \\
&&\Psi^{e}_{1,\pm} = e^{\pm i\phi_{\mathrm n}/\sqrt{2}}\,  e^{i3\phi_\rho/\sqrt{6}}\,, \quad\quad q=1,\quad h= 1\,,\nonumber
\end{eqnarray}
where we have indicated the corresponding electric charge, $q$, and the scaling dimension $h$. 
The $(l,m)$ indices in $\Psi^{qp/e}_{l,m}$ are quantum numbers of the neutral modes, classified according to representations of $SU(2)_1$ algebra \cite{KFP}. The operators $\Psi^{qp}_{l,m}$ are responsible for exchange of the quasiparticles between the bulk and the edge of the dot, while the operators $ \Psi^{e\dagger}_{1,\pm}$ create an electron (which is injected from the leads)  at the edge.

The state of the dot may be characterized by three numbers $(N,q,l)$: the integer net number of electrons, $N$, which have tunneled to the edge from the lead(s), the charge, $q$, which is  moved to the edge from the bulk (equals to  an integer times one third), and the number of neutral (zero mode) excitations at the edge (we will consider $l=0,\pm 1$). The energy of the dot is thus given by
\begin{equation}
E(N,q,l)=E_c\,q^2 +\frac{ v_\rho}{R}\, (N+q-\tilde V_g)^2 + \frac{ v_{\mathrm n}}{3 R}\, l^2\,,
                                                                            \label{eq:energy}
\end{equation}
where $E_c=e^2/2C$ is the bulk charging energy, $\tilde V_g = C_g V_g/e$ is dimensionless gate voltage and $v_{\rho/{\mathrm n}}$ is the velocity of the  charge/neutral edge mode, $v_{\mathrm n}<v_\rho$. Hereafter we shall assume that the bulk charging energy and neutral mode gap is smaller than the kinetic energy of the charge mode $E_c, { v_{\mathrm n}}/{3 R} < {v_\rho}/{R}$. This is the case, if the linear size of the dot $2\pi R$ is larger than its distance to the gate, $d$: the dot capacitance is $C\propto L^2/d$. In writing Eq.~(\ref{eq:energy}) we have also assumed that the gate is much more strongly  coupled to the edge, than to the bulk of the dot. This is indeed the case in most experimental setups.

\begin{figure}
\begin{center}
  \includegraphics[width=9cm]{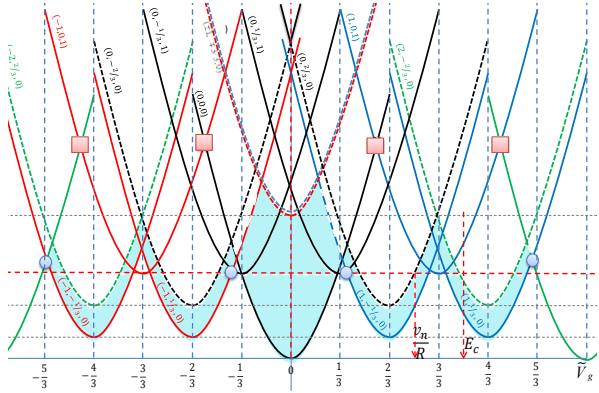}
  \vspace{-1cm}
\caption{(Color online)  Energy spectra and quantum numbers in CB $\nu=2/3$ quantum dot. Parabolas, plotted vs. the (normalized) gate voltage, are characterized by three quantum numbers, $(N,q,l)$: the total number of electrons,  the  charge moved from the edge from to the bulk, and  the number of neutral zero mode excitations. Each value 
of $N$ is represented by a parabola centered at that $N$ and same-color satellite parabolas, which are shifted horizontally  and vertically. Coulomb blockade peaks are obtained at the crossing of different color (different $N$) parabolas. The CB peaks occur at (blue) circles.   The shaded area between the ground state curve and the first excited state corresponding to a different value of $N$ (different color parabola), describes Coulomb  diamonds (cf. Fig. \ref{fig3}). } \label{fig2}
\end{center}
\end{figure}

\begin{figure}
\vspace{-1.5cm}
\hspace{-1cm}
\begin{center}
  \includegraphics[width=9cm]{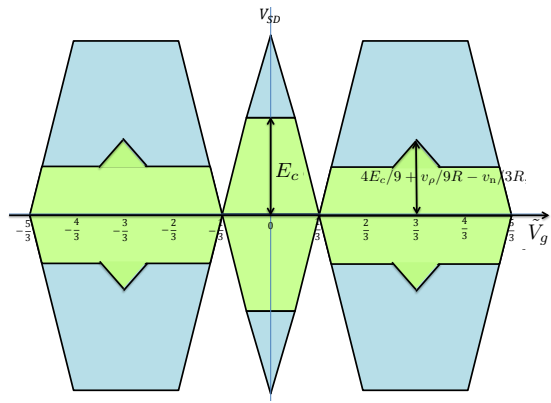}
  \vspace{-1cm}
\caption{(Color online) Schematic CB diamonds for  $\nu=2/3$. The green regions denote regions with zero conductivity \cite{Comment2}, while the purple
areas represent regions of parameters where the conductance is finite, but small since charge transfer involves both lead--edge and edge--bulk processes. Note the energy (voltage) scales that depend on $v_{{\mathrm n}}/L$.}
\label{fig3}
\end{center}
\end{figure}

At some gate voltage, hereafter taken to be $\tilde V_g=0$, the ground state of the dot is labeled as $(0,0,0)$. Upon varying the gate voltage, $\tilde V_g$, the energy of the dot increases quadratically as $v_\rho\tilde V_g^2/R$ -- the  (black) parabola centered at $\tilde V_g=0$   in Fig.~\ref{fig2}-- denoted as $(0,0,0)$. One may expect that the ground state  follows this parabola until it becomes energetically favorable to move an electron from the lead to the edge, utilizing, say,  the operator  $\Psi^{e\dagger}_{1,+}$. This would bring the dot to the state $(1,0,1)$ - the (blue) parabola centered at $\tilde V_g=3/3$. The crossing between these two parabolas, marked by a (purple) square,  takes place at $\tilde V_g \gtrsim 1/2$ at a rather high energy of the order ${v_\rho}/{R}$. According to this logic \cite{Stern}, the next time electron tunnels it is with the aid of $\Psi^{e\dagger}_{1,-}$ at $\tilde V_g\lesssim 3/2$, bringing the system to the state $(2,0,0)$ - the (green) parabola centered at $\tilde V_g=6/3$. This state is equivalent to the initial $(0,0,0)$, with two added electrons. As a result the CB periodicity  corresponds to  {\em two} electrons, as first realized in Ref.~\cite{Meir}. The odd-even peak spacing modulation  scales as $v_{\mathrm n}/v_\rho < 1$, and is expected to be small \cite{Stern}.

We note, however, that quasiparticle exchange between the edge and the bulk qualitatively modifies the sequence outlined above.
Starting again from the $(0,0,0)$ state and varying the gate voltage voltage $\tilde V_g$, one arrives at the point where moving one quasiparticle
from  the bulk to the edge becomes advantageous. 
Removing a quasiparticle from the edge  is achieved with the aid  of operators such as  $\Psi^{qp\dagger}_{1,+}$. The latter brings the system to the state $(0,1/3,1)$. We note that this change of the system's ground state involving edge-bulk charge rearrangement can be observed using a sensitive charge detector. 
 Since in all experiments the gate voltage is varied adiabatically, the system is bound to follow its lowest energy state, which
involves such quasiparticles exchange. Following the $(0,1/3,1)$  state, the system arrives at  the point (a (blue) circle in Fig.~\ref{fig2}),
where an electron enters from the lead ($N$ is changed by $1$), while a charge $2/3$ quasiparticle returns into the bulk. The corresponding composite operator
$ \Psi^{qp}_{0,0} \Psi^{e\dagger}_{1,-}$ brings the system into the state $(1,-1/3,0)$. It does have an additional electron, but contrary to the scenario outlined above, does not involve neutral mode excitation. As a result, the only energy to pay is for a $1/3$ charge in the bulk, which is the smallest energy scale here. The system then proceeds through a sequence of states which differ  by quasiparticle rearrangements between the bulk and the edge.  First to the $(1,0,1)$ state with the aid of $\Psi^{qp\dagger}_{1,+}$, then to $(1,1/3,0)$ state with the aid of $\Psi^{qp\dagger}_{1,-}$. Only then it becomes advantageous to bring in yet another electron from the lead and return $2/3$ quasiparticle to the bulk  with the composite operator $\Psi^{qp}_{0,0}\Psi^{e\dagger}_{1,+} $, making a transition to the state $(2,-1/3,1)$. Finally moving another quasiparticle from the bulk with $\Psi^{qp\dagger}_{1,-}$, the cycle is complete, and the system arrives at the  $(2,0,0)$ state.

Once again, the CB period is two electrons, in agreement with Refs.~\cite{Meir,Stern}. The CB peaks however are positioned at  markedly different locations vis-a-vis the gate voltage variation. 
In the limit $E_c\ll {v_\rho}/{R}$ the latter  tend to $\tilde V_g\approx 1/3, 5/3, 7/3, 11/3, \ldots$. Hence, one  approaches the {\em universal} $2:1$ ratio of odd--even CB peak spacings. One may argue that, since the electron tunneling is now accompanied by edge-bulk rearrangement, the peak conductance scales as a higher power of temperature and is rather small. While this is a valid argument, notice that,  upon adiabatic tuning of the gate voltage, the system has no choice, but to follow its true ground state. Thus the only CB peaks which may be ever observed are those where the {\em  ground state}
changes its number of electrons by one. Although the source-drain conductance happens to be small at this location, no other CB peaks could be reached in an adiabatic  measurement.

Possibly more interesting than the   $2:1$ peak spacing,  are the characteristics of the  CB diamonds. The latter are observed in the non-linear regime upon the application of a finite source--drain voltage $V_{SD}$. For a fixed gate voltage away from a CB peak, say $\tilde V_g=0$, the linear source--drain conductance is exceedingly
small.  It stays small until  $V_{SD}$ reaches the energy of an excited state which differs from the
ground state by the addition of one electron. For $\tilde V_g=0$ the simplest of such excited state $(1,0,\pm 1)$ has a rather high energy  $(v_\rho+ v_{\mathrm n}/3)/ R$. However, there are other excited states contributing to nonlinear source--drain conductance at substantially smaller voltage $V_{SD}$. In addition to electron tunneling from the lead, they involve bulk-edge exchange of quasiparticles. Examples of such states are
$(1,  -1/3, 0)$, $(1,-2/3,\pm 1)$ and $(1,-3/3,0)$. The latter (with  energy $E_c$) is the lowest excited state.
The energy difference between the $N=1$ (dashed blue) (or $N=-1$ (dashed red)) lowest excited state  and the $N=0$ (black) ground state (shown in blue shade) corresponds to the range of  $V_{SD}$ with vanishing differential conductance (cf. Fig.~\ref{fig3}), i.e. a CB diamond. For $N$-even ground states, i.e. between narrowly spaced CB peaks, the source-drain width of the diamond is about the bulk charging energy, $E_c$.
Let us turn now to the regions of $N$-odd ground state, e.g. $N=1$ (blue). The current carrying excitations are now states with $N=0$ and $N=2$.
It easy to see that the lowest energy ones are $(0,2/3,0)$ (dashed black) and $(2,-2/3,0)$ (dashed green). Their excitation energy in the middle of the diamond, i.e. $\tilde V_g=3/3$, is $4E_c/9+v_\rho/9R -v_{\mathrm n}/3R$, typically smaller than 
in $N$-even ground states. The width of the small triangles centered around $\tilde V_g=\pm 3/3$ is 
$\delta \tilde V_g=RE_c/(3v_\rho)+1/3 -v_{\mathrm n}/v_\rho$. One can therefore determine all three energy scales $E_c$, $v_\rho/R$ and $ v_{\mathrm n}/R$ from the shape of CB diamonds.


\begin{figure}
\hspace{-1cm}
\begin{center}
  \includegraphics[width=9cm]{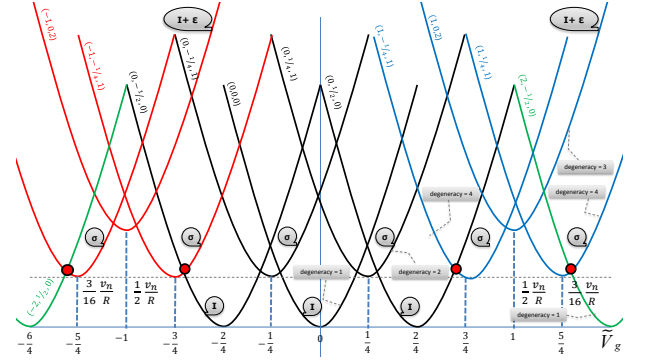}
\caption{(Color online) Energy spectra, quantum numbers $(N,q,l)$ and Klein factors $(\sigma,I,\epsilon)$ for $\nu=5/2$.  Degeneracy of various states are indicated. The CB peaks occur at (red) circles. }
\label{fig4}
\end{center}
\end{figure}

We now briefly outline details and results of a similar analysis for the anti-Pfaffian (APF) state \cite{Grosfeld,Rosenow+08,Rosenow+09,Parsa}, believed to be a candidate for the ground state of the  $5/2$ fraction \cite{Lee,Levin}. Due to its non-Abelian nature the electron and quasiparticle operators carry 
Klein factors, $(\sigma,I,\epsilon)$, \cite{Kitaev,Law} with  non-trivial fusion rules.  These factors select possible sequence of transitions which an initial ground-state may undergo upon changing the gate voltage along with the degeneracy of corresponding states \cite{Bonderson}.  The corresponding set of the operators $\Psi^{qp/e}_{l,m}$ is
\begin{eqnarray}
&&\Psi^{qp}_{1,\pm 1} = \sigma\cdot e^{\pm i\phi_{\mathrm n}/\sqrt{4}}\,  e^{i\phi_\rho/\sqrt{8}}\,, \quad\quad q={1\over 4},\quad h={1\over 4}\,;\nonumber \\
&&\Psi^{qp}_{0,0} =  I\cdot e^{i\phi_\rho/\sqrt{2}}\,, \quad\quad q={2\over 4},\quad h={1\over 4}\,; \label{eq:quasiparticles}  \\
&&\Psi^{e}_{2,\pm 2} = I\cdot e^{\pm i\phi_{\mathrm n}}\,  e^{i\phi_\rho\sqrt{2}}\,, \quad\quad q=1,\quad h= {3\over 2}\,,\nonumber\\
&&\Psi^{e}_{2,0} = \epsilon\cdot   e^{i\phi_\rho\sqrt{2}}\,, \quad\quad q=1,\quad h= {3\over 2}\,,\nonumber
\end{eqnarray}
supplemented by the fusion rules: $\epsilon\times\epsilon=I$, $\epsilon\times \sigma=\sigma$ and $\sigma\times\sigma = 
I+\epsilon$.  

Upon varying the gate voltage the initial ground state, labeled as $(0,0,0)$ and $I$, gives way to the internally rearranged $\sigma$ double-degenerate state $(0,1/4,\pm 1)$, Fig.~\ref{fig4}, obtained by transferring quasiparticle to the bulk with the operator $\Psi^{qp}_{1,\pm}$. The latter is succeeded by non-degenerate $I$ state $(0,1/2,0)$, obtained by moving yet another quasiparticle with the operator $\Psi^{qp}_{1,\mp}$. Only then it is energetically favorable to bring an external electron and move another quasiparticle with the help of either $\Psi^{e}_{2,0}\Psi^{qp}_{1,\pm}$, or $\Psi^{e}_{2,\pm 2}\Psi^{qp}_{1,\mp}$, leading to 4 times degenerate $\sigma$ state. This marks the position of the first CB peak. The cycle is then completed by successively moving two more quasiparticles to the bulk by going through $(1,0,2)$ and $(1,1/4,1)$. The latter then 
allows to bring the second electron and to move back a quasiparticle, arriving at $I$ state $(2, -1/4,0)$. This completes the 
two electron periodicity cycle.  As seen in Fig.~\ref{fig4}, the distance between successive CB peaks approaches a $3:1$ ratio.  Similarly to the $2/3$ case, the corresponding CB diamonds are narrow in the source drain direction in between the distant CB peaks and wider in between the closely spaced peaks.

%

To conclude, we have analyzed CB in quantum  dots with composite structure of the edge modes. We have shown that the presence of neutral modes leads to distinct features visible both in linear and non-linear conductance measurements. In particular, the sequence of  linear response CB peaks exhibits double periodicity with the universal peak distance ratio.
The shape and size of non-linear CB diamonds, on the other hand, offer a way to measure edge modes velocities for both charge and neutral edge excitations.

\begin{acknowledgments}
 We thank Eytan Grosfeld, Moty Heiblum, Yigal Meir and Ady Stern for illuminating discussions and Parsa Bonderson for very useful correspondence. AK was supported by NSF grant DMR1306734.  YG was supported by grants from the German-Israel Foundation, from BSF, and DFG grant RO 2247/8-1.
\end{acknowledgments}


\begin{thebibliography}{99}
\bibitem{Comment2} In the present discussion we have neglected cotunneling dominated transport, which is small and involves high order terms of electron and quasiparticle operators. This will be discussed elsewhere.
\end{thebibliography}
\end{document}